\documentclass[%
 reprint,
 amsmath,amssymb,
 aps,
]{revtex4-2}

\usepackage{graphicx}
\usepackage{dcolumn}
\usepackage{bm}
\usepackage{multirow}

\begin{document}

\preprint{APS/123-QED}

\title{Heavy Axions Can Disrupt $\gamma$-ray Bursts}

\author{Oindrila Ghosh}
    \email{oindrila.ghosh@fysik.su.se}
\author{Sunniva Jacobsen}%

\author{Tim Linden}%
 \email{trlinden@fysik.su.se}
\affiliation{%
 Stockholm University and The Oskar Klein Centre for Cosmoparticle Physics, Alba Nova, 10691 Stockholm, Sweden\\
}%

\date{\today}

\begin{abstract}

Axion-like particles (ALPs) can be produced in the hot dense plasma of fireballs that develop in the initial stage of $\gamma$-ray burst (GRB) outflows. They can transport an enormous amount of energy away from the jet by propagating out of the fireball. The photons produced by the eventual decay of such ALPs do not reach a sufficient density to re-thermalize through pair production, preventing fireball re-emergence. Thus, the production of heavy ALPs disrupts the fireball and dims GRBs, allowing bright GRB observations to strongly constrain the existence of heavy ALPs. By adding ALP interactions to existing models of GRB fireballs, we set competitive bounds on the ALP-photon coupling down to $g_{a \gamma \gamma} \sim 4 \times 10^{-12}~{\mathrm{GeV}^{-1}}$ for ALPs in the mass range of 200~MeV -- 5~GeV. 

\end{abstract}


\maketitle

\section{Introduction}

$\gamma$-ray bursts (GRBs), the most luminous objects in the visible universe, are thought to be powered by a central engine that is likely either a core-collapse supernova in the case of long GRBs, or the merger of compact objects in the case of short GRBs (sGRBs) \citep{piran1999gamma}. Their central engine launches a bipolar relativistic ejecta that produces a bright, highly variable, non-thermal prompt emission, followed by a long-lasting multi-wavelength afterglow which arises when the jet interacts with the circumburst medium \citep{salafia2022structure}. 

The most widely employed energy dissipation mechanism is the Blandford-Znajek (BZ) mechanism, which magnetically powers the jet by extracting the rotational energy of a spinning black hole in the presence of a large-scale poloidal field. This field is sustained by an accretion disk, leading to the creation of a rarefied funnel that defines the region where the jet eventually forms \citep{blandford1977electromagnetic}. Alternatively, neutrino-antineutrino annihilation \citep{popham1999hyperaccreting, ruffert1998gamma} and proto-magnetar central engines \citep{thompson2004magnetar, metzger2011protomagnetar} can also power jet formation. The luminosity generated via the BZ mechanism depends on the gravitational mass of the spinning black hole $M_{BH}$, the black hole spin parameter $a$, and the magnitude of the poloidal magnetic field $B$ following the relation $L_{BZ} \propto B^2 M_{BH}^2 a^2 \approx 10^{49}-10^{51}~\mathrm{erg/s}$, which is similar to the observed luminosity of GRBs.

    \begin{figure}
        \centering
        \includegraphics[width=0.44\textwidth]{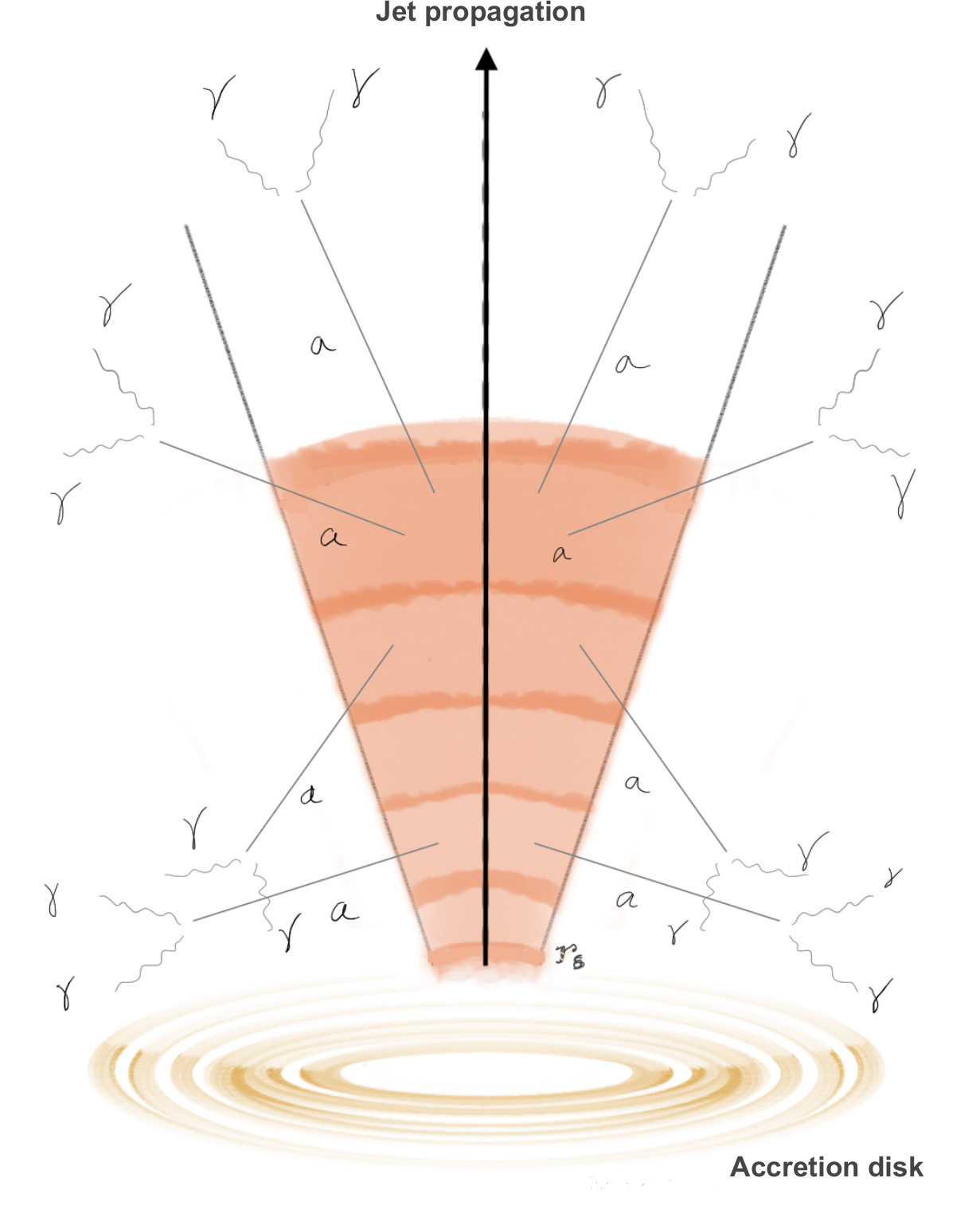}
        \caption{A schematic depicting a GRB outflow that is launched at the gravitational radius $r_s$ and is propagating within a cone. When ALPs are produced in the fireball, they isotropically emerge from it and decay into photons outside the source, disrupting the fireball (end of the orange region).}
        \label{fig:jetcartoon}
    \end{figure}

 The very-high-energy $\gamma$-rays from the central engine interact with the dense surrounding medium and produce prompt $e^{\pm}$ pairs which cascade to lower energy photons. The pair production opacity of the prompt GRB photons is enormous, implying a thermal spectrum and an underlying optically thick source. However, the observed non-thermal spectrum suggests that GRBs are optically thin. Fireballs were first proposed to reconcile this ``compactness problem" using the concept of a relativistic collimated outflow from a GRB \citep{goodman1986gamma, paczynski1986gamma}. The rapid variability timescale of sGRBs ($\delta t \sim 10~\mathrm{ms}$) requires that the source be extremely compact, with $R_i<c \delta T \approx 3000 \mathrm{~km}$. Allowing the source of radiation to move towards the observer at a relativistic speed, \emph{i.e.} a relativistic outflow, resolves this discrepancy because the upper limit on the source size is significantly larger $R_e<\gamma^2 c \delta t$ for a highly relativistic outflow $\gamma \gg 1$.

Even though fireballs radiate as a thermal blackbody, the observed GRB spectrum is nonthermal. This is because the electrons radiating in synchrotron have a power-law distribution due to the Fermi acceleration, associated with the strong forward and reverse shocks that occur when the ejecta decelerates upon encountering the external medium~\citep{rees1992relativistic}. Furthermore, this explains the radiative inefficiency of GRBs, since a substantial fraction of the bulk kinetic energy is re-thermalized in the presence of the external mass. The radius at which the fireball becomes optically thin is the photospheric radius, which is characterized by the emission region and is typically set to be $r_{\mathrm{ph}} = 10^{16}-10^{17}~\mathrm{cm}$. Because the outflow has undergone significant adiabatic cooling by this stage, the thermal contribution of the fireball to the GRB spectrum is subdominant.

Axion-like particles (ALPs) can be thermally produced in a multitude of astrophysical environments, in particular within dense relativistic or non-relativistic plasmas \citep{raffelt1996stars}, ranging from the early universe \citep{langhoff2022irreducible} to the cores of astrophysical objects. In the latter case, their weak interactions with the surrounding material allow them to efficiently transport energy out of astrophysical systems. For heavy axions, the constraints on the Standard Model (SM) couplings are weaker due to the high energy scale needed to produce ALPs. In addition to cosmological bounds on ALPs from irreducible freeze-in contribution \citep{langhoff2022irreducible}, Big Bang Nucleosynthesis (BBN) \citep{depta2021updated} and the effective number of neutrino species at BBN \citep{cadamuro2012cosmological}, supernovae and other stellar cooling bounds provide the strongest constraints on heavy axions \citep{giannotti2016cool, muller2023investigating, carenza2020constraints, calore2021supernova, caputo2022radiative, caputo2022low}. However, more recently, leading bounds have also emerged from merging compact objects and their associated GRB emission~\citep{fiorillo2022axions, dev2024first, diamond2024multimessenger, diamond2023axion, reynoso2017production}. 

In the very hot plasma contained within the GRB fireball, ALPs can be produced in great abundance and subsequently decay into photons and electrons. At this stage, the fireball is a pure photon-lepton fireball which is a blackbody at a very high temperature. Baryons are swept up as the fireball expands through the circumburst medium. The evolution of such a blackbody is governed by pair production, pair annihilation, and bremsstrahlung. The production of heavy ALPs that strongly couple to this photon-lepton plasma provide a unique opportunity to probe ALP physics. A schematic of the scenario we present here is depicted in Fig. \ref{fig:jetcartoon}.


In this article, we explore various ALP production mechanisms that can occur in a photon-lepton fireball and demonstrate that heavy ALPs can disrupt the fireball at a very early stage of its evolution. In Section \ref{sec:fireball}, we discuss the classic fireball model in the absence of axions. In Section \ref{sec:processes}, we compute the rates of ALP-based processes in this region, comparing them with the SM processes that govern the evolution of SM particles in the fireball. We also discuss how particle decay and gravitational trapping can affect ALP escape from the fireball. Finally, in Section \ref{sec:discuss}, we discuss the results and show our constraints in the ALP-photon parameter space, before concluding in Section \ref{sec:concl}.

\section{Fireball structure and properties}
\label{sec:fireball}


\subsection{Evolution of the fireball}

The various stages in the evolution of a fireball are analogous to that of the early universe \citep{piran1999gamma}. First, a pure radiation fireball is formed at a relativistic temperature that is sufficient to kinematically allow pair production. Once the pairs are created, the opacity of the fireball increases and the radiation cannot escape. This photon-lepton fireball fluid expands and cools as $T \propto r^{-1}$. Pairs annihilate as soon as this temperature drops below the pair-production threshold and the number of pairs is significantly reduced by $T \sim 20~\mathrm{keV}$ causing the pair plasma to become optically thin and allowing the photons to escape. The fireball is accelerated during this phase with $\gamma \propto r$ and expands relativistically. 

Over time, baryonic matter begins to be swept up, and the free electrons associated with the baryons further increase the opacity and delay the escape of radiation until the fireball has cooled down even further. In the matter-dominated phase, the baryons swept up by the fireball undergo acceleration which spends the bulk kinetic energy of the fireball during this stage, eventually leading to the asymptotic coasting of the plasma with a constant Lorentz factor. This phase continues until the conversion of the internal energy into the kinetic energy is complete \citep{paczynski1990super, shemi1990appearance}. These phases can be approximately divided as follows:

\begin{itemize}
    \item (a) A Pure Radiation Fireball ($ r \lesssim 10^7-10^8~\mathrm{cm}$): The effect of the baryons is negligible and the fireball is radiation-dominated. This describes a pure photon-lepton fireball. 
    \item (b) Electron-dominated Fireball ($ 10^8~\mathrm{cm} < r < 10^9~\mathrm{cm}$): In later stages, free electrons associated with baryons are prevalent but the fireball is still radiation-dominated and loses most of its energy as radiation.
    \item (c) Relativistic Baryonic Fireball ($ r \sim 10^9~\mathrm{cm}$): Before becoming optically thin, the fireball is matter-dominated after re-thermalizing the swept up mass. The internal energy is further converted into the bulk kinetic energy of the baryons which become nearly relativistic, leading to a significant deceleration of the fireball.
    \item (d) Newtonian Fireball ($r > 10^9~\mathrm{cm}$): In the Newtonian regime, the expansion is no longer relativistic. This scenario is distinct from the relativistic outflow in the initial stages of fast transients and exemplifies the physical pictures of objects such as supernova explosions.
\end{itemize}

\subsection{Geometry of the Fireball}

Initially, fireballs were assumed to be isotropically expanding shells of ejecta with a spherical geometry. However, it is commonly observed that the radiative efficiencies of the brightest GRBs are low, which implies an extremely large explosion energy associated with the bursts \citep{meszaros2019gamma}. A natural solution to this problem is to envision a collimation of the outflow. 

For slower ejecta that arise from tidal heating in the accretion disk, this can occur through the launching of a wind, either originating from the magnetic reconnection events between the magnetospheres of the merging binary systems or pair creation through neutrino-antineutrino annihilation, which eventually forms a jet \citep{meszaros1992tidal}. For mergers of compact objects, a gravitational focusing of the pairs can collimate the outflow into a jet aligned perpendicular to the accretion disk \citep{meszaros1992high}. There are various configurations of the jet model when the flow is confined within a cone: which can be one-zone and characterized by a homogeneous medium, or can be multizone, \emph{e.g.} a structured GRB jet with an inner spine and a surrounding sheath \citep{salafia2018jet}. 

For the purpose of this work, we limit ourselves to a one-zone collimated outflow that is released within the collimated flow defined by the jet opening angle $\Delta \theta \approx 10^{-1}$ \citep{escorial2023jet}. The volume of the conical shell is $\Delta V = \pi \Delta \theta^2 r^2 \Delta r$ where $r$ is the distance from the central engine and $\Delta r$ is the thickness of the shell. This is consistent with the fact that for highly relativistic outflows, the jet is beamed at the angle $\Delta \theta \sim \gamma^{-1}$. In order to obtain the intrinsic true luminosity of a GRB jet from its observed isotropic-equivalent luminosity, a correction involving the beaming factor $f_b \sim (1 - \Delta \theta) \approx (\Delta \theta^2/2)$ needs to be applied such that $L_{\mathrm{intr}} \approx (\Delta \theta^2/2) L_{\mathrm{iso}} \approx L_{\mathrm{iso}}/\gamma^2$. For long and short GRBs, the best-fit jet opening angles are $\Delta \theta \approx 4^{\circ}~\mathrm{and}~16^{\circ}$ respectively \citep{frail2001beaming, fong2015decade}.

For sGRBs that are associated with binary compact object mergers, radiation escapes along the rotation axis in a funnel-like fashion, leading to a strong departure from sphericity. Despite the complexity of the geometry of this system, the highly relativistic nature of the outflow allows us to consider the jet to be part of a spherical shell where the scaling relations of a homogeneous radiative fireball provide a good approximation \citep{piran1999gamma}.

\subsection{Physical Parameters}

The steady-state outflow from a GRB can be described by Flammang's equations, which can be solved to obtain \citep{paczynski1986gamma, goodman1986gamma} the following scaling relations in the purely radiative initial stage of the fireball:

\begin{equation}
    \gamma=r / r_0
\end{equation}

\begin{equation}
    T=T_0 r_0 / r
\end{equation}

The above scaling laws are characteristic of a fluid undergoing uniform comoving expansion, similar to an expanding universe. 

For a short GRB produced from a binary merger, where each of the compact objects has masses $M_{\mathrm{rem}} \sim \mathcal{O}(1.5 M_{\odot})$, the gravitational radius of the remnant with mass of $\sim \mathcal{O}(3 M_{\odot})$ is 

\begin{equation}
    r_s = 8.86 \times 10^{5} \left(\frac{M_{\mathrm{rem}}}{3M_{\odot}} \right)^{1/2}~\mathrm{cm}.
\end{equation}

This is the radius at which the first fireball is launched from the central engine and the outflow starts to move at an accelerated speed. The Lorentz factor for this motion is specified above. Therefore, at $r = r_s$, the fireball is an extremely dense compact blob of plasma reaching temperatures $T_s \sim \mathcal{O}(100~\mathrm{MeV})$. In this work, we primarily focus on this earliest stage of the GRB fireball, where most of the ALP processes take place. We will not discuss the stage in the life of the GRB jet where baryons start having an effect, beyond $r \sim 10^8~\mathrm{cm}$. The distance from the remnant at which the fireball is launched varies based on the model of GRB formation that is adopted. In most of the relevant literature, the collimated outflow is described from the distance scale of $r = 10^7-10^8~\text{cm}$. However, this does not necessarily reflect the beginning of the fireball, in particular for smaller remnants, but rather the distance scale relevant to the fireball starting to interact with the circumburst medium and evolve from its purely radiative initial stage. 

\section{Relevant Processes}
\label{sec:processes}

The rates of SM and ALP processes depend on the plasma temperature $T(r)$ at a given fireball radius $r$.

\subsection{Standard Model processes}

The SM processes that maintain the thermodynamic equilibrium of the fireball are pair production, annihilation, and Bremsstrahlung \citep{svensson1982electron}. For unpolarized photons, the pair-production cross-section is \citep{jauch2012theory, svensson1982electron} 

\begin{equation}
\sigma_{\gamma \gamma \rightarrow e^{+} e^{-}}=r_e^2 \pi\left(\frac{m_e}{E_{\gamma}}\right)^2\left(\ln \frac{2 E_{\gamma}}{m_e}-1\right),
\end{equation}

\noindent where the classical electron radius is $r_{\mathrm{e}}= \alpha/m_{\mathrm{e}}$ with $\alpha = 1/137$ as the fine-structure constant, and $E_\gamma$ is the incoming photon energy. Pair production turns on above the threshold fireball temperature of $T= 10 m_e$ and the corresponding rate is given by:

\begin{equation}
\begin{aligned}
&\Gamma_{\gamma \gamma \rightarrow e^{+} e^{-}}= \begin{cases}0, & T<10 m_e \\ n_\gamma (T) \cdot \sigma_{\gamma \gamma \rightarrow e^{+} e^{-}}  & T \geq 10 m_e\end{cases}.\\
\end{aligned}
\end{equation}

The pair annihilation rate depends on the number density of $e^{\pm}$ in the fireball $n_e(T)$, which we assume to be in equipartition with the photon density $n_\gamma (T)$.  The thermally-averaged pair annihilation cross-section in a hot relativistic one-temperature thermal plasma is estimated to be \citep{svensson1982pair}

\begin{equation}
\langle\overline{\sigma v}\rangle= r_e^2 \frac{\pi}{2 \mathcal{\vartheta}^2} \ln 2 \eta_E \vartheta \quad(\vartheta \gg 1)
\end{equation}

\noindent where the dimensionless temperature parameter is \mbox{$\vartheta = T / m_e$}, which enters the pair annihilation rate as

\begin{equation}
    \Gamma_{e^{+} e^{-} \rightarrow \gamma \gamma} = n_{e}(T) \langle\overline{\sigma v}\rangle_{e^{+} e^{-}  \rightarrow \gamma \gamma}
\end{equation}

\noindent where $\gamma_E$ is the Euler-Mascheroni constant. In the relativistic temperature limit $\vartheta \gg 1$, the Bremmstrahlung emissivity can be written as

\begin{equation}
\mathcal{W}_{ee \rightarrow ee \gamma}(\vartheta)=24 n_e(T)^2 r_e^2 \alpha \vartheta[\ln (2 \vartheta / \gamma_E)+5 / 4].
\end{equation}

Since the average energy of the photon emitted via Bremsstrahlung $e e \rightarrow e e \gamma$ is approximately $3 T$ in a thermal bath, the Bremsstrahlung rate is then \citep{alexanian1968photon} 

\begin{equation}
\Gamma_{ee \rightarrow ee \gamma}(\vartheta) = \mathcal{W}_{ee \rightarrow ee \gamma}(\vartheta)/3T .
\end{equation}

The SM process rates are shown in Fig. \ref{fig:allprocesses-1} in salmon, yellow, and orange, which are significantly larger than the rates for ALP processes. The photon and electron number densities and the rates for the above SM processes are sensitive to any changes in the fireball temperature as it expands and evolves apart from ALP production. However, we do not solve the Boltzmann equation for each species since the relative number of photons and electrons are maintained even when slightly out of equilibrium. In the next section, we intensify our efforts to understand the impact of the ALP processes on the total luminosity and power of the fireball.

\subsection{ALP production processes}

The major ALP production processes in the leptonic fireball can be broadly classified into two categories. First, we have the 2$\rightarrow$1 processes:

\begin{itemize}
    \item [] a. Inverse decay involving photons (also known as photon fusion): $\gamma \gamma \rightarrow a$
    \item [] b. Inverse decay involving $e^{\pm}$: $e^{-} e^{+} \rightarrow a$
\end{itemize}

\noindent as well as the 2$\rightarrow$2 processes: 

\begin{itemize}
    \item [] a. Photon conversion: $e^{ \pm} \gamma \rightarrow e^{ \pm} a$ 
    \item [] b. Fermion ($e^{\pm}$) annihilation: $e^{-} e^{+} \rightarrow \gamma a$
\end{itemize}

 For $2 \rightarrow 1$ processes, we can calculate the rate as

\begin{equation}
\begin{aligned}
\Gamma_{\text{a},\text{2}\rightarrow\text{1}}(T) & = \frac{\left|\mathcal{M}_{1+2 \rightarrow a}\right|^2}{32 \pi^3} \\ & 
\times \int_{m_a}^{\infty} d E_a f_a^{\mathrm{eq}}\left(\beta p_a+2 T \ln \left[\frac{1 \mp e^{-E_{+} / T}}{1 \mp e^{-E_{-} / T}}\right]\right)
\end{aligned}
\end{equation}

\noindent where $E_{ \pm}=\left(E_a \pm \beta p_a\right) / 2$ with $\beta=\sqrt{1-4 m_1^2 / m_a^2}$ where $m_1$ is the mass of the first incoming particle. For $\gamma$ inverse decay, the squared amplitude reads 

\begin{equation}
\sum\left|\mathcal{M}_{\gamma \gamma \rightarrow a}\right|^2=\frac{1}{2} g_{a \gamma \gamma}^2 m_a^2\left(m_a^2-4 m_\gamma^2\right),
\end{equation}

\noindent where the thermal mass of the photon depends on the temperature of the photon bath, which in our case refers to the temperature of the fireball $m_\gamma(T) \simeq e T / 3 \simeq T / 10$ \citep{raffelt1996stars}. The squared amplitude for $e$ inverse decay can be written as

\begin{equation}
\sum\left|\mathcal{M}_{e^{-} e^{+} \rightarrow a}\right|^2=2 g_{a e e}^2 m_a^2.
\end{equation}

For $2 \rightarrow 2$ processes, the rate can be computed as
\begin{equation}
\begin{aligned}
\Gamma_{a,2 \rightarrow 2}(T) & = \frac{g_1 g_2 T}{32 \pi^4} \\ &
\times \int_{s_{\min }}^{\infty} d s \lambda\left(s, m_1^2, m_2^2\right) \frac{K_1(\sqrt{s} / T)}{\sqrt{s}} \sigma_{12 \rightarrow 3 a}(s)
\end{aligned}
\end{equation}

\noindent with $d^4 P=2 \pi \sqrt{P_0^2-s} d P_0 d s$, $s=P^2 \in\left[s_{\min }, \infty\right)$, and $s_{\min }=\left(m_1+m_2\right)^2$, noting that

\begin{equation}
\int d \Pi_1 d \Pi_2 \delta^4\left(P-p_1-p_2\right)=\frac{1}{(2 \pi)^6} \frac{4 \pi \lambda^{1 / 2}\left(P^2, m_1^2, m_2^2\right)}{8 P^2}
\end{equation}

\noindent where $\lambda(a, b, c) \equiv a^2+b^2+c^2-2 a b-2 b c-2 c a$. For photon conversion, the cross-section can be written as:

\begin{equation}
\begin{aligned}
\sigma_{e^{ \pm} \gamma \rightarrow e^{ \pm} a}(s) & = \frac{\alpha g_{a \gamma \gamma}^2}{32 s^2}\bigg[2\left(2 s^2-2 m_a^2 s+m_a^4\right) \ln \left(\frac{s-m_a^2} {m_\gamma^2}\right) \\
&-7 s^2+10 m_a^2 s-5 m_a^4\bigg] \\
& +\frac{\alpha g_{a e e}^2}{8 s^3}\bigg[2\left(2 s^2-2 m_a^2 s+m_a^4\right) \ln \left(\frac{s}{m_e^2}\right) \\
&-3 s^2+10 m_a^2 s-7 m_a^4\bigg] \\
& -\frac{\alpha g_{a \gamma \gamma} g_{a e e} m_e}{8 s^3\left(s-m_a^2+m_e^2\right)} \\
& \times \bigg[2\left(s^3+m_a^6\right) \ln \left(\frac{\left(s-m_a^2\right)^2}{\left(s+m_a^2\right) m_e^2}\right) \\
&-3\left(s+m_a^2\right)\left(s-m_a^2\right)^2\bigg].
\end{aligned}
\end{equation}

Finally, the cross-section for $e^{\pm}$ annihilation is given by:

\begin{equation}
\begin{aligned}
\sigma_{e^{-} e^{+} \rightarrow \gamma a}(s) & = \frac{\alpha g_{a \gamma \gamma}^2}{24 \beta}\left(1-\frac{m_a^2}{s}\right)^3\left(1+\frac{2 m_e^2}{s}\right)  \\ & +\frac{\alpha g_{a e e}^2}{2 s^2\left(s-m_a^2\right) \beta^2} \\ & \times \left[\left(s^2-4 m_e^2 m_a^2+m_a^4\right) \ln \left(\frac{1+\beta}{1-\beta}\right) -2 \beta m_a^2 s\right] \\
&-\frac{\alpha g_{a \gamma \gamma} g_{a e e} m_e}{2 s \beta^2}\left(1-\frac{m_a^2}{s}\right)^2 \ln \left(\frac{1+\beta}{1-\beta}\right)
\end{aligned}
\end{equation}

\noindent where $\beta=\sqrt{1-4 m_e^2 / s}$.

\begin{figure*}[htbp]
        \centering
        \includegraphics[width=1.1\textwidth]{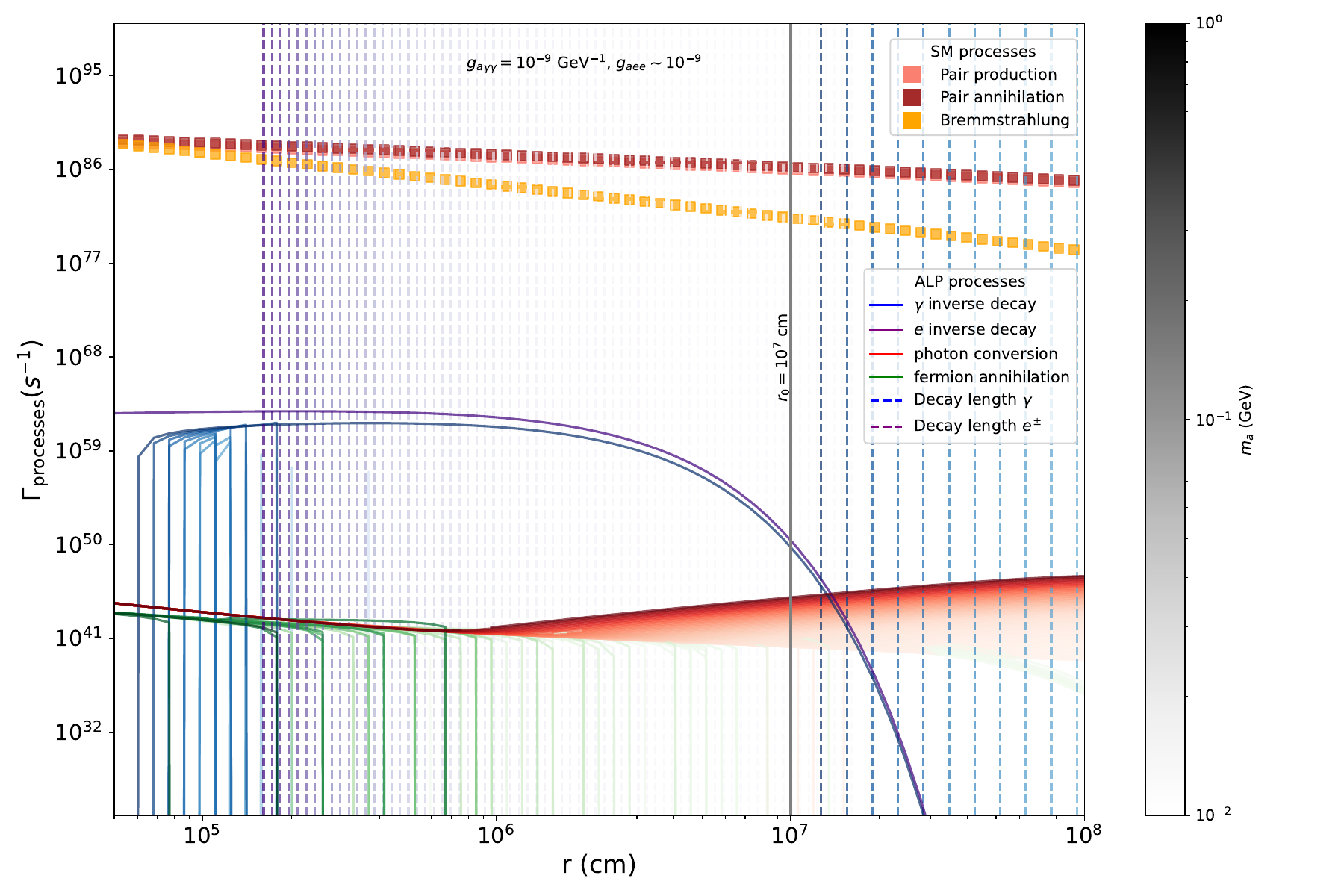}
        \caption{Total rates for all relevant processes obtained by integrating over the entire fireball, presented as the $\Gamma_{\text{processes}}$ at each radius $r$(cm). presented in this paper. The rates are divided into two broad categories. Standard Model processes: pair production (salmon), Bremsstrahlung (orange) and pair annihilation (brown), are shown as squares. ALP processes: photon inverse decay (blue), electron inverse decay (purple), photon conversion (red), and Fermion annihilation (green) are shown as solid lines for ALP-photon couplings of $g_{a\gamma\gamma} = 10^{-9}~\text{GeV}^{-1}$ and ALP-electron couplings of $g_{aee} = 10^{-9}$. For ALP processes, lighter shades correspond to lighter ALPs and deeper shades signify heavier ALPs, with the colorbar ranging from 10 MeV to 1 GeV in logarithmic scale.  The major backreactions are ALP decays into photons and electrons. The decay lengths for ALPs into photons (electrons) are shown in blue (purple) dashed lines, assuming a conservative $\gamma$-factor of 1.05 corresponding to the fireball expansion speed. In the case of $e$ inverse decay, the rates are relatively independent of the ALP mass, and all points overlap.}
        \label{fig:allprocesses-1}
    \end{figure*}

In Figure~\ref{fig:allprocesses-1}, we plot the rates of these processes against the distance from the central engine, and compute the decay length conservatively assuming an ALP Lorentz factor $\gamma_a = E_a/m_a = 1.05$ corresponding to the fireball expansion speed, which is significantly smaller than the smallest ALP Lorentz factor we calculate in Table \ref{tab:gamma_factors}. We have calculated the process rates assuming a thermal spectra of $e^{\pm}$ similar to \citep{langhoff2022irreducible, caputo2022low, muller2023investigating} and a photon effective mass of $m_{\gamma} \sim T/10$ for the plasma conditions in the fireball \citep{raffelt1996stars}. The fireball environment bears similarities to the early universe with its constituent relativistic plasma undergoing accelerated expansion. The primary backreactions are ALP decays into photons and $e^{\pm}$ pairs, the rate of which depends strongly on the ALP mass.

From Fig. \ref{fig:allprocesses-1}, we immediately see that the 2$\rightarrow$1 ALP processes of photon (blue) and electron (purple) inverse decays have the highest rates among the ALP (production) processes. The 2$\rightarrow$2 ALP processes, photon conversion (red) and fermion annihilation (green), are not as efficient even at very high temperatures. We also note that SM processes, pair production (salmon), Bremsstrahlung (orange) and pair annihilation (brown) all have rates that are higher than the ALP processes for the ALP-photon and ALP-electron couplings we consider, implying that the state of the fireball is still governed by SM processes that are in equilibrium even when ALP processes have high rates. 

We also note that ALP decay into electrons is very efficient for large ALP masses. If the ALP is heavy and has a large electron coupling, it will decay primarily inside the fireball without being able to escape, as can be seen from the purple dashed lines. Thus, heavy ALPs with large electron couplings cannot disrupt the fireball. However, in the case of ALP-photon interactions, ALPs primarily decay outside the fireball for all relevant ALP masses, which ends up disrupting the fireball. Therefore, in this investigation, we concern ourselves with very heavy ALPs that only have photon couplings ($g_{aee} = 0$). 

\begin{figure*}[htbp]
        \centering
        \includegraphics[width=1.1\textwidth]{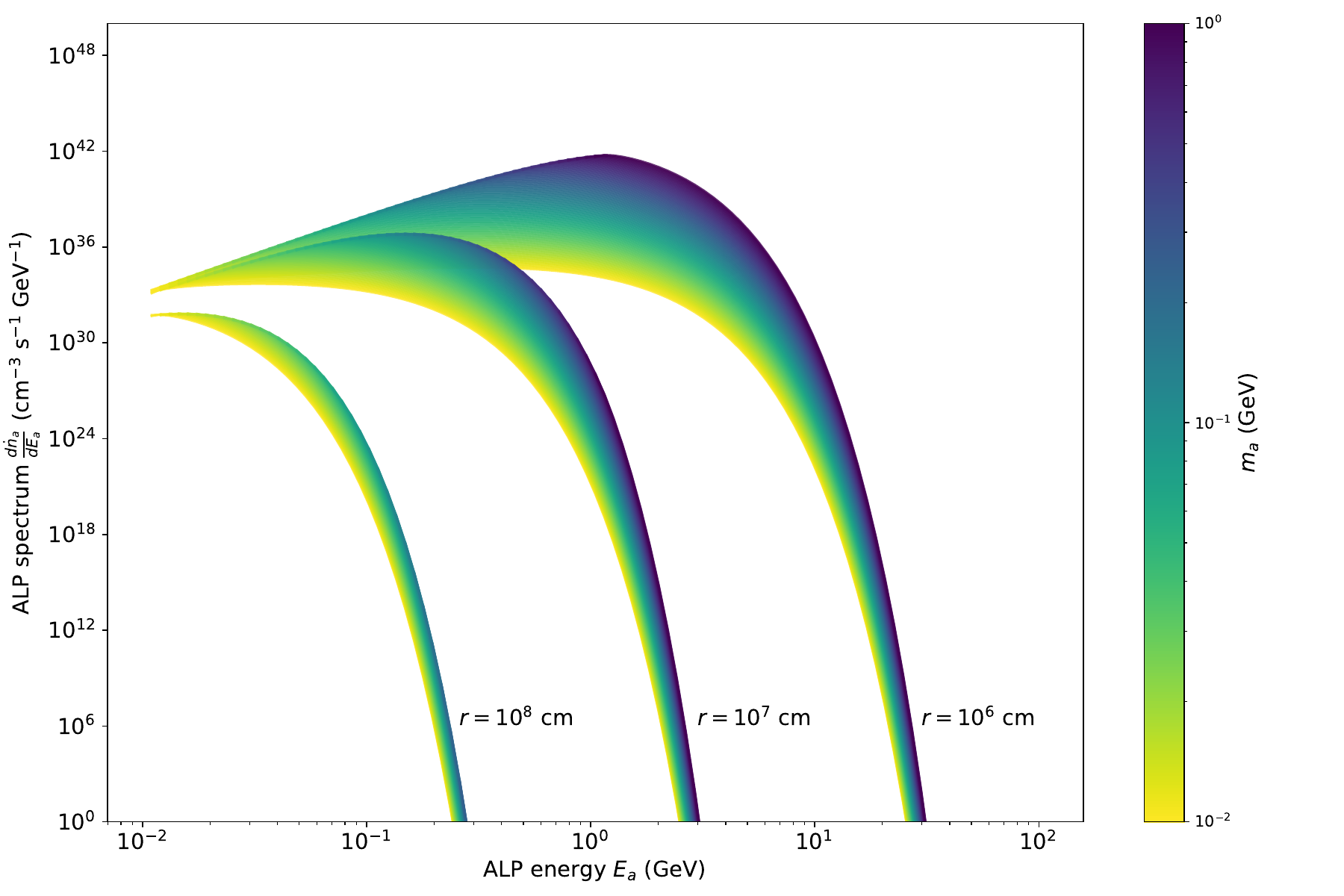}
        \caption{Spectrum of ALPs produced per unit volume through photon fusion calculated at different fireball radii of $r = 10^6$, $10^7$, and $10^{8}$ cm. The colorbar represents the ALP mass, $m_a$ going from 10 MeV (light yellow) to 1 GeV (deep purple)
        \label{fig:alpspectra}}
    \end{figure*}

In this case the primary production mechanisms are photon fusion, also known as the inverse decay of photons, shown in blue and the Primakoff production, also known as photon conversion (PC), shown in red. The photon conversion (Primakoff production) of ALPs is subdominant for ALPs throughout the ALP mass range considered here \citep{ferreira2022strong, lucente2022constraining}, as can also be seen from Fig. \ref{fig:allprocesses-1} . The spectra of ALPs produced thermally through photon fusion between energy $E_a$ and $E_a + dE_a$ can be written as:

\begin{equation}
\frac{d \dot{n}_a}{d E_a}(T)=\frac{g_{a \gamma \gamma}^2}{128 \pi^3} m_a^4 p\left(1-\frac{4 \omega_{\mathrm{pl}}^2}{m_a^2}\right)^{3 / 2} e^{-E_a / T}
\label{eq:ALPspectra}
\end{equation}

where $n_a$ is the number density of the axions produced with mass $m_a$, energy $E_a$, and a photon coupling $g_{a\gamma \gamma}$ inside the fireball. Here $p = \sqrt{E_a^2 - m_a^2}$ and $\omega_{\mathrm{pl}} << m_a$ for our mass range of interest, so we can ignore the term within the parentheses. As seen from Fig. \ref{fig:allprocesses-1} and Eq. \ref{eq:ALPspectra}, the rate and the spectrum of the ALPs produced depend strongly on the temperature and therefore the fireball radius as $T(r)$. We show the spectrum of the ALP produced per unit volume at three distinct lengthscales, $R = 10^6$, $10^7$, and $10^{8}$ cm in Fig. \ref{fig:alpspectra}.

In Fig. \ref{fig:alpspectra}, we see that the production of heavier ALPs through $\gamma$ inverse decay or photon fusion peaks at higher temperatures, which are achieved at smaller fireball radii. Photons in the hot thermal plasma of the GRB fireball follow a Planck distribution, and it is the highest-energy photons in their tail that have energies above the kinematic threshold needed to produce the ALPs ($E_a \geq m_a$) and thus participate in the photon fusion more than lower-energy photons. This has to do with the fact that the phase space scales as $p_a = \sqrt{E_a^2 - m_a^2}$ and $\gamma \gamma \rightarrow a$ sets $E_a = 2 E_{\gamma}$. Thus, at high temperatures when $E_a \gg m_a$ , the phase space scales as the incoming photon energy $p_a \propto E_{\gamma}$ leading to a higher rate of ALP production.

The mean Lorentz factor of the ALPs can be calculated by using the ALP spectrum:

\begin{equation}
\langle\gamma_a\rangle (m_a, T)=\frac{\int \frac{E_a}{m_a} \cdot \frac{d \dot{n}_a}{d E_a} (m_a, T) d E_a (m_a, T)}{\int \frac{d \dot{n}_a}{d E_a}(m_a, T) d E_a}.
\end{equation}

We show the average Lorentz factors for the ALPs that are produced in Table \ref{tab:gamma_factors} .

\begin{table}
\centering
\begin{tabular}{|c|c|c|c|}
\hline
$r \, $ & $ T \, (\text{GeV}) $ & $ m_a \, (\text{GeV}) $ & $ \langle \gamma_a \rangle $ \\
\hline
\multirow{5}{*}{$r_s(2M_{\odot}) = 5.93 \times 10^5 \text{cm}$} & \multirow{5}{*}{$0.506$} & $0.01$ & $100.984308$ \\
 &  & $0.08$ & $12.935285$ \\
 &  & $0.2$ & $5.551709$ \\
 &  & $0.5$ & $2.713063$ \\
 &  & $1$ & $1.817522$ \\
\hline
\multirow{5}{*}{$r_s(3M_{\odot}) = 8.90 \times 10^5 \text{cm}$} & \multirow{5}{*}{$0.337$} & $0.01$ & $67.377468$ \\
 &  & $0.08$ & $8.808573$ \\
 &  & $0.2$ & $3.956192$ \\
 &  & $0.5$ & $2.111106$ \\
 &  & $1$ & $1.529511$ \\
\hline
\multirow{5}{*}{$r_s(4M_{\odot}) = 1.19 \times 10^6 \text{cm}$} & \multirow{5}{*}{$0.253$} & $0.01$ & $50.584786$ \\
 &  & $0.08$ & $6.765675$ \\
 &  & $0.2$ & $3.174323$ \\
 &  & $0.5$ & $1.817522$ \\
 &  & $1$ & $1.385863$ \\
\hline
\multirow{2}{*}{$r_c = 3.00 \times 10^7 \text{cm}$} & \multirow{2}{*}{$0.01$} & $0.01$ & $2.692544$ \\
 &  & $0.08$ & $1.115430$ \\
\hline
\end{tabular}
\caption{Average Lorentz factors of ALPs for gravitational radii of $2M_{\odot}$, $3M_{\odot}$, $4M_{\odot}$ and a critical radius $R_c$ at which the jet is formed, considering ALP masses of $m_a = 10~\text{MeV},~80~\text{MeV},~200~\text{MeV},~500~\text{MeV},~1~\text{GeV}$.}
\label{tab:gamma_factors}
\end{table}

\subsection{Gravitational trapping}
\label{subsec:gravtrap}

Heavier ALPs are more likely to be produced at low velocities, making them more succeptible to gravitational trapping. This can prevent them from exiting the fireball before decaying. Because these ALPs decay (and then re-thermalize) within the fireball, they do not contribute to the disruption of the fireball. In order to account for this correction, we compute the ALP luminosity as

\begin{equation}
\begin{aligned}
L_a &= \pi \Delta \theta^2 \int_{r_s}^{r_c} d r \, r^2 \int_{m_a}^{\infty} d E_a \, E_a \, \frac{d \dot{n}_a}{d E_a} \\ &
\times \Theta \left( E_a - m_a - \frac{2 G M m_a}{r c^2} \right).
\label{eq:gravlum}
\end{aligned}
\end{equation}

The lower limit of the integral in Eq. \ref{eq:gravlum} corresponds to the gravitational radius of the remnant, $r_s$, i.e., where the fireball is launched. The upper limit corresponds to the radius of the fireball where it starts becoming matter-dominated, giving rise to the jet. Even the most conservative model for the outflow speed at this distance scale applies a Lorentz boost of $\gamma_{\mathrm{ejecta}} \geq 1$ corresponding to $v_{\text{ejecta}} \approx 0.1c$\citep{lithwick2001lower}.  We can thus estimate this critical radius to be

\begin{equation}
    r_c \approx  3 \times 10^7~\mathrm{cm} \left (\frac{v_{\mathrm{ejecta}} }{0.1c}\right)\left(\frac{\delta t }{10~\text{ms}}\right),
\end{equation}

\noindent where $\delta t \sim 10~\mathrm{ms}$ is the typical variability timescale for sGRBs. Most of the ALP production takes place within this distance scale. We note in passing that the correction owing to gravitational trapping does not affect our results significantly.


We have assumed a conservative ALP Lorentz factor $\gamma_{a,\mathrm{min}} \approx 1.05$ in the central engine frame. This corresponds to the fireball expansion speed of $v_{\text{exp}}\sim 0.3c$ \citep{villar2017combined}, defining a limiting case for which ALPs have reached the threshold speed to leave the fireball. From Table \ref{tab:gamma_factors}, we see that the average Lorentz factor of the ALP produced exceeds this limiting case even for a fireball radius of $r_c = 3 \times 10^7$ cm corresponding to a cooler fireball temperature of 10 MeV. However, the detailed modelling of this late stage lies outside the scope of this paper, because in this work we only consider a pure photon-lepton radiative fireball. Beyond $r_c = 3 \times 10^7$~cm, the fireball starts becoming matter-dominated and we plan to explore this scenario in our future work.

\subsection{Trapping by decay}
\vspace{-0.4cm}
In order to determine whether photophilic ALPs produced in the fireball contribute to decreasing the fireball luminosity, we need to determine whether these ALPs decay inside or outside the rapidly expanding fireball. If ALPs decay inside the fireball, they re-thermalize, and do not contribute to the dissolution of the fireball. The relevant decay lengthscale for photophilic ALPs is:

\begin{equation}
\lambda_{a \rightarrow \gamma \gamma}=\frac{64 \pi}{g_{a \gamma}^2 m_a^4} \sqrt{E_a^2 - m_a^2}.
\end{equation}

Considering this, we apply a decay-corrected luminosity, which is estimated as:
\vspace{-0.3cm}
\begin{equation}
L_a= \pi \Delta \theta^2 \int_{r_s}^{R_c} d r \, r^2 \int_{m_a}^{\infty} d E_a \, E_a \, \frac{d \dot{n}_a}{d E_a} e^{-r/\lambda_{a \rightarrow \gamma \gamma}}
\end{equation}

In order for the ALPs to escape the fireball before decaying, they should have a critical speed of $\gamma_{a, \text{cr}} \approx \gamma_{\text{exp}} = 1.05$, while the average ALP speed at a given radius, as shown in Table \ref{tab:gamma_factors}, exceeds it $\langle \gamma_{a} \rangle \geq \gamma_{a, \text{cr}}$.

If ALPs in the mass range 100 MeV--1 GeV are ejected from the fireball, their decay back into the photon field produces a photon flux too rarefied to thermalize via pair creation. ALP-mediated fireballs in GRBs produced via photon fusion and subsequent decay cannot form for $m_a \geq 60~\mathrm{MeV}$ \citep{diamond2023axion, diamond2024multimessenger}. This implies that the classical fireball launched by the BZ mechanism is disrupted.

\section{Results and discussion}
\label{sec:discuss}

\subsection{ALP constraint from GRB luminosities}

Due to the extremely efficient ALP production during the early stages of the compact fireball, where temperatures can reach close to $T \sim 10^{12}~K$, a huge population of ALPs will remove power from the fireball. If these ALPs can propagate out of the fireball at a rate faster than the fireball expansion rate without decaying inside, an enormous amount of energy is removed through ALP transport, thus disrupting the fireball and subsequently the relativistic outflow. 

Therefore, the strongest constraints on heavy ALPs come from the observed luminosity of sGRBs, by requiring that the ALP luminosity $L_{a}$ cannot exceed the intrinsic GRB luminosity. We estimate the GRB luminosity by adjusting the observed isotropic-equivalent luminosities of GRBs in $\gamma$-rays $L_{iso}$ by the beaming factor $\Delta \theta^2/2$, such that $L_a \leq \Delta \theta^2 L_{iso} / 2 \approx 10^{50}~\mathrm{erg/s}$. 

\begin{figure}
        \centering
        \includegraphics[width=0.5\textwidth, trim=0 0 2 2, clip]{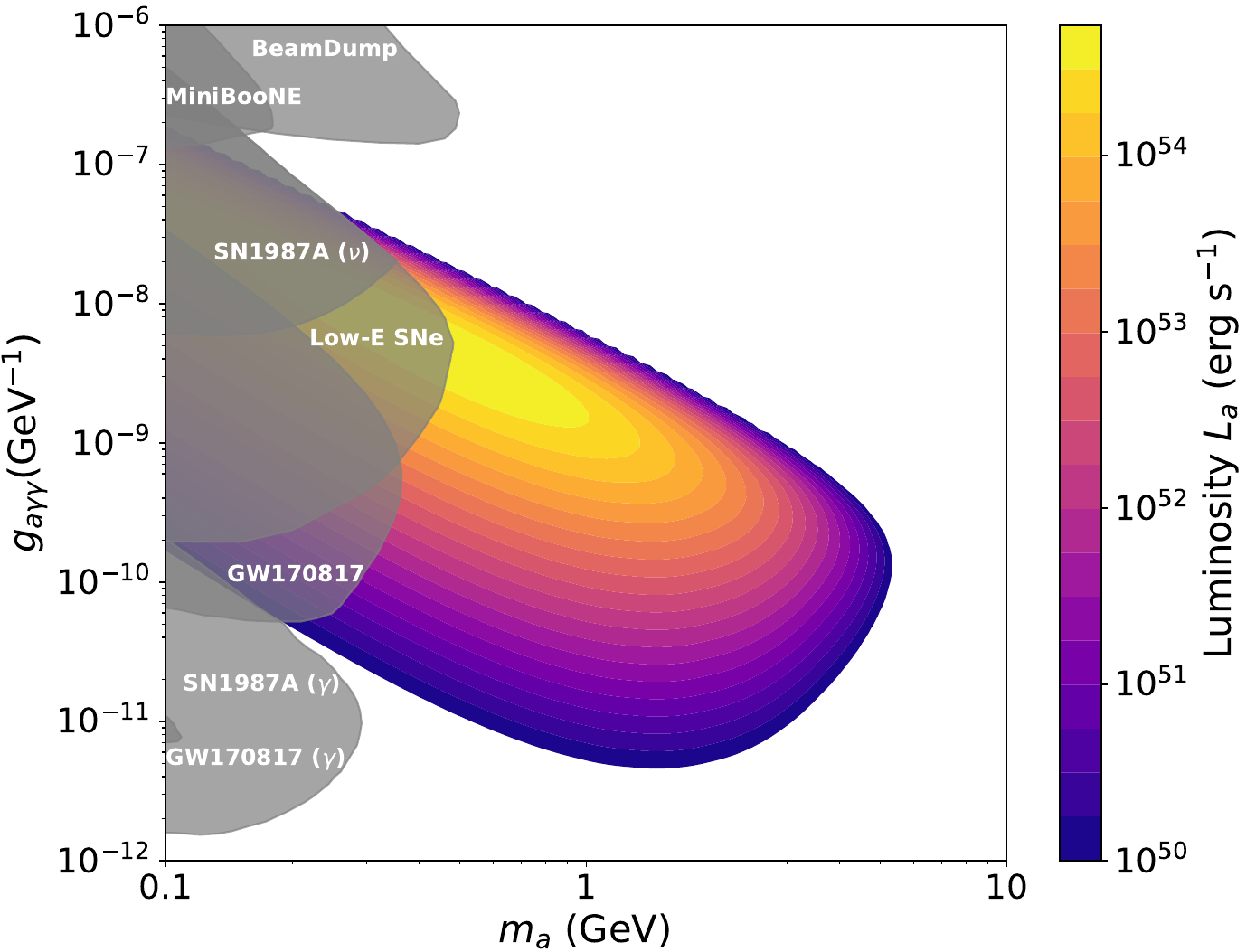}
        \caption{Luminosity exclusion contours in the ALP-photon parameter space for a remnant mass of $M_{\mathrm{rem}} = 3 M_{\odot}$. The grey contours show existing constraints from various astrophysical observations.}
        \label{fig:agconstraint}
    \end{figure}

In Figure~\ref{fig:agconstraint},  we show the resulting exclusion region in the $m_a$-$g_{a \gamma \gamma}$ plane for ALPs that will disrupt the GRB, with different colored regions representing the luminosity that ALPs will remove from the GRB, in units of $\mathrm{erg}~\mathrm{s}^{-1}$, which we plot on a logarithmic scale. The lower bound of these limits is set by the fact that smaller couplings will produce a smaller ALP luminosity. The upper bound is set by the fact that large couplings will cause ALPs to decay back into photons before leaving the fireball. The upper mass limit is set by the maximum temperature in the GRB fireball. The gray regions show existing bounds. Our results show leading limits on ALP mass and ALP-photon coupling, reaching $g_{a \gamma \gamma} \sim 10^{-12}~\mathrm{GeV}^{-1}$ and GeV-scale ALP masses.

\subsection{Limitations and applicability}

This work relates to leptonic sources associated with GRBs and provides physical arguments supporting a process where heavy ALPs can be produced in the hot dense plasma of the initial stage of a GRB fireball, and subsequently carry an enormous amount of energy out of the developing fireball, disrupting or dimming the source. We have shown that the decay photons from the heavy ALPs we consider here ($m_a >$ 100 MeV) typically decay far from the fireball center, and thus cannot further re-thermalize to form a fireball again. However, the decay photons can marginally contribute to isotropic $\gamma$-ray backgrounds or lead to an extended emission, which is likely to be too faint to detect \citep{ackermann2015spectrum}. In fact, the evolution of the GRB fireball presents us with the opportunity to probe ALP masses across the spectrum, from the heavier GeV-scale to lighter MeV-scale ALP masses, through their production in the thermal plasma as it expands and cools. The constraints derived from the later stages of the fireball are similar to those outlined in \citep{diamond2023axion, dev2024first}.

Previous studies focusing on ALP production in the merger remnant \citep{dev2024first} and accretion disk \citep{reynoso2017production}, as well as ALP-mediated fireballs \citep{diamond2023axion}  in the transient, have explored potential signals arising from the decay of the ALP produced, through the interaction of the decay products with ambient radiation or magnetic field. On the contrary, our strong leading ALP-photon bounds rely exclusively on a lack of detection, i.e., if these heavy ALPs were to exist, they would be produced in copious amounts in the initial stages of the hot fireball and transport a sufficient amount of energy away from the source to dim or extinguish the prompt emission that is observed from GRB jets. 

After baryon loading begins at a later stage of the fireball, nucleon-nucleon interactions start playing an important role in ALP production. In addition, there are interesting observational consequence for ALPs that have both electron and nucleon couplings. This is because the cascade photons arising from decay pairs will lead to a time delay in the photon flux due to scatterings with the intergalactic magnetic field. This process can be indirectly constrained using GRB secondary cascade photons, thus leading to a nonlinear feedback in the calculation of IGMF from transients. This is a topic of our ongoing work. The full impact of an axio-electric and axio-nuclear cascade in transients have been explored in the context of accretion disks and remnants, but we plan to investigate their impact in the GRB jet structure in future work. 

\section{Conclusions}
\label{sec:concl}

In this article, we have explored energy losses from GRBs during the initial stages of its evolution, which are driven by ALP production and escape from the hot dense fireball plasma. We considered a purely photophilic ALP with masses in the range $\mathcal{O}$(10 MeV)---$\mathcal{O}$(1 GeV), for which the leading production mechanism is photon fusion $\gamma \gamma \rightarrow a$. These ALPs decay into photons far outside the fireball, thus disrupting its evolution. Once the ALPs have emerged from the jet plasma and decayed into photons, the decay photons cannot form a photon field sufficiently dense for them to re-thermalize and recreate a fireball outside of the jet plasma. Thus, we show that observations of prompt photons from GRBs produce the strongest constraint on heavy ALPs in the mass range of $m_a \sim 200~\mathrm{MeV} - 5~\mathrm{GeV}$, where they are sensitive to ALP-photon coupling down to $g_{a\gamma \gamma} \sim 4\times10^{-12}~\mathrm{GeV}^{-1}$. To achieve this sensitivity, we require that the ALP luminosity cannot exceed a canonical isotropic GRB luminosity of $L_{\text{iso}} \sim 10^{52}$ erg/s corrected for relativistic beaming as $L_{a} \lesssim \Delta \theta^2 L_{\text{iso}} /2  \sim 10^{50} \text{erg/s}$. More specific constraints can be drawn based on the least to the most luminous observed GRBs, such as the GRB 201015A with a peculiarly low luminosity of $L_{\text{iso}} \sim 10^{50}$ erg/s \citep{markwardt2020grb} and the spectacularly luminous GRB 221009A which has an inferred isotropic luminosity of $L_{\text{iso}} \sim 10^{54}$ erg/s \citep{huang2022lhaaso}. 

We note that we have computed the rates of all SM and ALP processes occurring in the fireball plasma for ALPs with both photon and electron couplings. Even though leptophilic ALPs are produced more copiously at higher temperatures occurring at the initial stages of the fireball, they decay into $e^{\pm}$ pairs before emerging from the fireball for a significant part of the parameter space, leaving little to no observable difference from a GRB fireball evolving without ALP processes. However, potential constraints may be obtained on heavy leptophilic ALPs produced in the GRB fireball for extremely small values of $g_{aee}$ through their contribution to various cosmic radiation backgrounds.

\acknowledgements{}
We acknowledge the AxionLimits repository, maintained by Ciaran O'Hare, which we used to plot the existing constraints in Fig. \ref{fig:agconstraint}. OG is supported by the European Research Council under Grant No. 742104 and by the Swedish Research Council (VR) under the grants 2018-03641 and 2019-02337. SJ acknowledge support by the Vetenskapsradet (Swedish Research Council) through contract No. 638-2013-8993 and the Oskar Klein Centre for Cosmoparticle Physics. TL acknowledges support by the Swedish Research Council under contract 2022-04283.

\nocite{*}

\bibliography{alpgrb}

\end{document}
%